\documentclass[pre,12pt]{revtex4}

\usepackage{amsmath}    % need for subequations
\usepackage{graphicx}   % need for figures
\usepackage{verbatim}   % useful for program listings
\pdfoutput=1 

\begin{document}

\title{Simple algorithm for GCD of polynomials}
\author{Pasquale Nardone, Giorgio Sonnino}
\email{pasquale.nardone@ulb.be}
\affiliation{Physics Department, Universit\'e Libre de Bruxelles\\
50 av F. D. Roosevelt, 1050 Bruxelles, Belgium}
\date{\today}

\begin{abstract}
Based on the Bezout approach we propose a simple algorithm to determine the {\tt gcd} of two polynomials which doesn't need  division, like the Euclidean algorithm, or determinant calculations, like the Sylvester matrix algorithm. The algorithm needs only $n$ steps for polynomials of degree $n$. Formal manipulations give the discriminant or the resultant for any degree without needing division nor determinant calculation.

\end{abstract}

\maketitle

\section{Introduction}

There exist different approach to determine the greatest common divisor ({\tt gcd}) for two polynomials, most of them are based on Euclid algorithm~\cite{KK} or matrix manipulation~\cite{CC}~\cite{DD} or subresultant technics~\cite{AA}. All these methods requires long manipulations and calculations around $O(m^2)$ for polynomials of degree $m$. Bezout identity could be another approach.
 If $P^{(m)}(x)$ is a polynomial of degree $m$ and $Q^{(m)}(x)$ is a polynomial of degree at least $m$, the Bezout identity says that ${\tt gcd}(P^{(m)}(x),Q^{(m)}(x))=s(x)P^{(m)}(x)+t(x)Q^{(m)}(x)$ where $t(x)$ and $s(x)$ are polynomials of degree less then $m$. Finding $s(x)$ and $t(x)$ requires also $O(m^2)$ manipulations. If we know that $P^{(m)}(0)\neq0$ we propose here another approach which use only linear combination of $P^{(m)}(x)$ and $Q^{(m)}(x)$ and division by $x$ to decrease the degree of both polynomials by $1$.

\section{Algorithm}

Let's take two polynomials $P^{(m)}(x)$ and $Q^{(m)}(x)$:
$$
P^{(m)}(x)=\sum_{k=0}^{m}p_k^{(m)}x^k\quad;\quad Q^{(m)}(x)=\sum_{k=0}^{m}q_k^{(m)}x^k
$$
with $p_0^{(m)}\neq0$ and $p_m^{(m)}\neq0$. The corresponding list of coefficients are:
$$
p^{(m)}=\{p_0^{(m)},p_1^{(m)},\cdots,p_{m-1}^{(m)},p_m^{(m)}\}\quad;\quad q^{(m)}=\{q_0^{(m)},q_1^{(m)},\cdots,q_{m-1}^{(m)},q_m^{(m)}\}
$$
Let's define $\Delta_m=q_m^{(m)}p_0^{(m)}-p_m^{(m)}q_0^{(m)}$. If $\Delta_m\neq0$, we can
build two new polynomials of degree $m-1$ by cancelling the lowest degree term and the highest degree term:
\begin{equation}
\left\{\begin{aligned}
&P^{(m-1)}(x)={1\over x}\bigl(q^{(m)}_0 P^{(m)}(x)-p^{(m)}_0 Q^{(m)}(x)\bigr)\\
&Q^{(m-1)}(x)=q^{(m)}_m P^{(m)}(x)-p^{(m)}_m Q^{(m)}(x)
\end{aligned}\right.
\end{equation}
or in matrix notation:

\begin{equation}
\begin{pmatrix}
x\;P(x)\\
Q(x)\\
\end{pmatrix}_{m-1}=\begin{pmatrix} q_0^{(m)} & -p_0^{(m)} \\ q_m^{(m)} &-p_m^{(m)}\end{pmatrix}.\begin{pmatrix}
P(x)\\
Q(x)\\
\end{pmatrix}_{m}
\end{equation}
and the reverse:
\begin{equation}
\begin{pmatrix}
P(x)\\
Q(x)\\
\end{pmatrix}_{m}={1\over\Delta_m}\begin{pmatrix} -p_m^{(m)}& p_0^{(m)}\\ -q_m^{(m)} &q_0^{(m)} \end{pmatrix}.\begin{pmatrix}
x\;P(x)\\
Q(x)\\
\end{pmatrix}_{m-1}
\end{equation}
If $\Delta_m=0$ then we replace $Q^{(m)}(x)$ by ${\tilde Q}^{(m)}(x)$:
\begin{equation}
\left\{\begin{aligned}
&P^{(m)}(x)=P^{(m)}(x)\\
&{\tilde Q}^{(m)}(x)=x(p_0^{(m)}Q^{(m)}(x)-q_0^{(m)}P^{(m)}(x))
\end{aligned}\right.
\end{equation}

This correspond to the manipulation on the list of coefficients:

\begin{equation}
\text{if }\ \Delta_m\neq0\quad
\left\{
\begin{matrix}
p_k^{(m-1)}=q^{(m)}_0p^{(m)}_{k+1}-p^{(m)}_0 q^{(m)}_{k+1}\\
\\
q_k^{(m-1)}=q^{(m)}_m p^{(m)}_k-p^{(m)}_m q^{(m)}_k
\end{matrix}
\right.\quad k\in[0,m-1]
\end{equation}
or
\begin{equation}
\text{if }\ \Delta_m=0\quad
\left\{
\begin{matrix}
{\tilde q}_0^{(m)}=0\\
{\tilde q}_k^{(m)}=p_0^{(m)}q_{k-1}^{(m)}-q_0^{(m)}p_{k-1}^{(m)}
\end{matrix}
\right.\quad k\in[1,m]
\end{equation}
note that the new ${\tilde q}_1^{(m)}=0$. Note also that $p_{m-1}^{(m-1)}=-q_0^{(m-1)}=-\Delta_m$ and this will remains true at all iteration ending with $p_0^{(0)}=-q_0^{(0)}=-\Delta_1$.

So we have the same Bezout argument, (we know that $0$ is not a root of $P^{(m)}(x)$) the ${\tt gcd}(P^{(m)}(x),Q^{(m)}(x))$ must divide $P^{(m-1)}(x)$ and $Q^{(m-1)}(x)$ or  $P^{(m)}(x)$ and ${\tilde Q}^{(m)}(x)$. Repeating the iteration, it must divide $P^{(m-2)}(x)$ and $Q^{(m-2)}(x)$. If we reach a constant : $P^{(0)}(x)=p_0^{(0)}$ and $Q^{(0)}(x)=q_0^{(0)}=-p_0^{(0)}$ then ${\tt gcd}(P^{(m)}(x),Q^{(m)}(x))=1$. If we reach, at some stage $j$ of iteration, $P^{(m-j)}(x)=0$ or $Q^{(m-j)}(x)=0$ then the previous stage $j-1$ contains the ${\tt gcd}$.

When dealing with numbers the recurrence could gives large numbers so we can normalise the polynomials by some constant
\begin{equation}
\begin{aligned}
&P^{(m-1)}(x)={\alpha_{m-1}\over x}\bigl(q^{(m)}_0 P^{(m)}(x)-p^{(m)}_0 Q^{(m)}(x)\bigr)\\
&Q^{(m-1)}(x)=\beta_{m-1}\bigl(q^{(m)}_m P^{(m)}(x)-p^{(m)}_m Q^{(m)}(x)\bigr)
\end{aligned}
\end{equation}
choosing for example $\alpha$ and $\beta$ such that the sum of absolute value of the coefficients of  $P^{(m-1)}(x)$ and $Q^{(m-1)}(x)$ are $1$: $\alpha_{m-1}^{-1}=\sum_{k=0}^{m-1}|p_k^{(m-1)}|$, $\beta_{m-1}^{-1}=\sum_{k=0}^{m-1}|q_k^{(m-1)}|$, or that the maximum of the coefficients is always $1$: $\alpha_{m-1}^{-1}={\tt max}(p_k^{(m-1)})$, $\beta_{m-1}^{-1}={\tt max}(q_k^{(m-1)})$. 

For example if $P^{(8)}=x^8+x^6-3 x^4-3 x^3+8 x^2+2 x-5$ and $Q^{(8)}=3 x^6+5 x^4-4 x^2-9 x+21$, after 5 iterations we have to deal with numbers of order $10^{15}$, while using the sum of absolute value or the maximum we obtain after 8 iterations the result which prove that the polynomials are co-prime:
$$
\begin{aligned}
&P^{(8)}(x) \frac{\left(699 x^5-236877 x^4+8107 x^3-37558 x^2+11607 x+158088\right)}{130354 x^8}+\\
&-Q^{(8)}(x)\frac{ \left(233
   x^7-78959 x^6+2547 x^5+40120 x^4+1938 x^3+61457 x^2-3839 x-37640\right)}{130354 x^8}=1\\
&Q^{(8)}(x)\frac{  \left(7528 x^7-233 x^6+86487 x^5-2547 x^4-62704 x^3-24522 x^2-1233 x+18895\right)}{130354
   x^7}+\\
   &-P^{(8)}(x)\frac{ \left(22584 x^5-699 x^4+274517 x^3-8107 x^2+7446 x-79359\right)}{130354 x^7}=1
\end{aligned}
$$

In term of list manipulation we have:
$$
\text{if}\ \Delta_m\neq0\quad \begin{matrix}
p^{(m-1)}={\tt Drop}[{\tt First}[q^{(m)}]\;p^{(m)}-{\tt First}[p^{(m)}]\;q^{(m)},1]\\
q^{(m-1)}={\tt Drop}[{\tt Last}[q^{(m)}]\;p^{(m)}-{\tt Last}[p^{(m)}]\;q^{(m)},-1]
\end{matrix}
$$
where {\tt First[list]} and {\tt Last[list]} takes the first and the last element of the list respectively, while {\tt Drop[list,1]} and {\tt Drop[list,-1]} drop the first and the last element of the list respectively. If $\Delta_m=0$ then we know that $p_0^{(m)}q_m^{(m)}-q_0^{(m)}p_m^{(m)}=0$ so the list $p_0^{(m)}q^{(m)}-q_0^{(m)}p^{(m)}$ ends with $0$ so the list manipulation is :
$$
{\tilde q}^{(m)}={\tt RotateRight}[{\tt First}[p^{(m)}]q^{(m)}-{\tt First}[q^{(m)}]p^{(m)}]
$$
where {\tt RotateRight[list]} rotate the list to the right ({\tt RotateRight[\{a,b,c\}]=\{c,a,b\}}).

Repeating these steps decrease the degree of polynomials. So or we reach a constant, and reversing the process enables us to find a combinations of $P^{(m)}$ and $Q^{(m)}$ which gives a monomial $x^k$ and the polynomials are co-prime, or we reach a $0$-polynomial before  reaching the constant and $P^{(m)}(x)$, $Q^{(m)}(x)$ have a non trivial {\tt gcd}.

For example 
\begin{equation}
\begin{aligned}
&
\left\{\begin{aligned}
&P^{(8)}(x)=x^8-4 x^6+4 x^5-29 x^4+20 x^3+24 x^2+16 x+48\\
&Q^{(8)}(x)=x^8+3 x^7-7 x^4-21 x^3-6 x^2-18 x
\end{aligned}\right. \\
&\left\{\begin{aligned}
&p^{(8)}=\{48,16,24,20,-29,4,-4,0,1\}\\
&q^{(8)}=\{0,-18,-6,-21,-7,0,0,3,1\}
\end{aligned}\right.
\end{aligned}
\end{equation}
and let's use the ``max" normalisation. The first iteration says that ${\tt gcd}$ must divide $P^{(7)}(x)$ and $Q^{(7)}(x)$:
$$
P^{(7)}=-{1\over21x}Q^{(8)}\ \text{and}\ Q^{(7)}={1\over48}(P^{(8)}-Q^{(8)})
$$
$$
\left\{\begin{aligned}
&P^{(7)}(x)=-\frac{x^7}{21}-\frac{x^6}{7}+\frac{x^3}{3}+x^2+\frac{2x}{7}+\frac{6}{7}\\
&Q^{(7)}(x)=-\frac{x^7}{16}-\frac{x^6}{12}+\frac{x^5}{12}-\frac{11 x^4}{24}+\frac{41 x^3}{48}+\frac{5 x^2}{8}+\frac{17 x}{24}+1
\end{aligned}\right.
$$
then {\tt gcd} divide
$$
\left\{\begin{aligned}
&P^{(6)}(x)=\frac{x^6}{78}-\frac{2 x^5}{13}-\frac{2 x^4}{13}+\frac{11 x^3}{13}-\frac{67x^2}{78}+x-\frac{9}{13}\\
&Q^{(6)}(x)=\frac{x^6}{4}+\frac{x^5}{5}-\frac{11 x^4}{10}+x^3-\frac{33 x^2}{20}+\frac{4x}{5}-\frac{3}{10}
\end{aligned}\right.
$$
then {\tt gcd} divide
$$
\left\{\begin{aligned}
&P^{(5)}(x)=\frac{22 x^5}{57}+\frac{8 x^4}{19}-\frac{31 x^3}{19}+x^2-\frac{115 x}{57}+\frac{11}{19}\\
&Q^{(5)}(x)=-\frac{32x^5}{187}-\frac{19 x^4}{187}+\frac{155 x^3}{187}-\frac{151 x^2}{187}+x-\frac{12}{17}
\end{aligned}\right.
$$
etc.. finally {\tt gcd } divide
$$
\left\{\begin{aligned}
&P^{(3)}(x)=x^3+3 x^2+x+3\\
&Q^{(3)}(x)=\frac{x^3}{3}+x^2+\frac{x}{3}+1
\end{aligned}\right.
$$
the next step will give $Q^{(2)}(x)=0$ ($3Q^{(3)}(x)-P^{(3)}(x)=0$), with the last step:
$$
\left\{
\begin{aligned}
&P^{(2)}(x)=P^{(8)}(x) \left(\frac{88}{63 x^4}+\frac{50}{63 x^3}+\frac{229}{378 x^2}+\frac{143}{378 x}\right)+\\
&-Q^{(8)}(x)
   \left(-\frac{704}{189 x^5}-\frac{400}{189 x^4}+\frac{164}{189 x^3}-\frac{100}{189 x^2}+\frac{143}{378
   x}\right)=x^3+3 x^2+x+3\\
&Q^{(2)}(x)=P^{(8)}(x)  \left(-\frac{6}{x^3}+x-\frac{1}{x}\right)-Q^{(8)}(x)\left(\frac{16}{x^4}-\frac{8}{x^2}+x+\frac{4}{x}-3\right)=0
\end{aligned}\right.
$$
so we have ${\tt gcd}(P^{(8)}(x),Q^{(8)}(x))=x^3+3 x^2+x+3$

Doing the algorithm on formal polynomials gives automatically the resultant or the discriminant of $P^{(m)}(x)$ and  $Q^{(m)}(x)$.

For example for the {\tt gcd} of $P^{(m)}(x)$ and $P^{(m)}(x)'$ for formal polynomials (we always cancel the term $x^{m-1}$ by translation) we have:
$$
P^{(3)}(x)=a\;x^3+b\;x+c\quad Q^{(3)}(x)=3a\;x^2+b
$$
gives after 3 iterations the well known discrimant:
$$\begin{aligned}
&(9 a b c x+2b^3)P^{(3)}(x)-(3abc x^2+(2b^3+9ac^2)x+2b^2c)Q^{(3)}(x)=-a(4b^3+27ac^2)x^3\\
&3b(2bx-3c)P^{(3)}(x)+(9c^2+3bcx-2b^2x^2)Q^{(3)}(x)=(4b^3+27ac^2)x^2
\end{aligned}
$$
For the general polynomial of degree $4$:
$$
P^{(4)}=a\;x^4+b\;x^2+c\;x+d\quad Q^{(4)}=4a\;x^3+2b\;x+c
$$
in 5 iterations we have, if $3c^2-8bd\neq0$ the discriminant is~\cite{BB}
$$
{\tt disc}=256 a^2 d^3-128 a b^2 d^2+144 a b c^2 d-27 a c^4+16 b^4 d-4 b^3 c^2
$$
and
\begin{equation}
\begin{aligned}
&P^{(4)}\left(-4 c x \left(16 a d^2+12 b^2 d-3 b c^2\right)+8 x^2 \left(-16 a b d^2+6 a c^2 d+4 b^3 d-b^2c^2\right)+c^2 \left(9 c^2-32 b d\right)\right)+\\
&+Q^{(4)} \left(c x^2 \left(16 a d^2+12 b^2 d-3 b
c^2\right)-x \left(-64 a d^3+16 b^2 d^2-38 b c^2 d+9 c^4\right)+\right.\\
&\left.-2 x^3 \left(-16 a b d^2+6 a c^2 d+4
   b^3 d-b^2 c^2\right)-c d \left(9 c^2-32 b d\right)\right)={\tt disc}\;x^4
\end{aligned}
\end{equation}
and
\begin{equation}
\begin{aligned}
&P^{(4)} (2 c (3 b c^2 - 8 b^2 d + 32 a d^2) - 
     8 (-b^2 c^2 + 4 b^3 d + 6 a c^2 d - 16 a b d^2) x + 
     4 a c (9 c^2 - 32 b d) x^2) + \\
 &Q^{(4)} (2 d (-3 b c^2 + 8 b^2 d - 32 a d^2) + 
     2 c (-3 b c^2 + 10 b^2 d - 8 a d^2) x + \\
&+2 (-b^2 c^2 + 4 b^3 d + 6 a c^2 d - 16 a b d^2) x^2 - 
     a c (9 c^2 - 32 b d) x^3) =-{\tt disc}\; x^3
\end{aligned}
\end{equation}
if $3c^2-8bd=0$ the discriminant is
$$
{\tt disc}=27 a^2 c^4+18 a b^3 c^2+4 b^6
$$
and
\begin{equation}
\begin{aligned}
&P^{(4)} \left(-24 a b^3 c x^2+18 a b c^3-8 b^5 x\right)+\\
&+Q^{(4)} \left(\frac{3}{2} b c x \left(2 b^3-3 a c^2\right)+6a b^3 c x^3-\frac{27 a c^4}{4}+2 b^5 x^2\right)=-{\tt disc}\; x^3
\end{aligned}
\end{equation}
and
\begin{equation}
\begin{aligned}
&P^{(4)} (4 b^2 (2 b^3 + 9 a c^2) + 24 a b^3 c x + 72 a^2 b c^2 x^2) + \\
& Q^{(4)} (-{3\over2} b c (2 b^3 + 9 a c^2) - 2 b^2 (b^3 + 9 a c^2) x - 
    6 a b^3 c x^2 - 18 a^2 b c^2 x^3)={\tt disc}\; x^2
\end{aligned}
\end{equation}

A more formal case~\cite{BB} is:
$$
P^{(m)}(x)=x^m+a\;x+b\quad ;\quad P^{(m)}(x)'=m\;x^{m-1}+a
$$
so we have successively:
\begin{equation}
k\in[0,m]\ :\ p_k^{(m)}=b\;\delta^k_0+a\;\delta^k_1+\delta^k_m\ \text{and}\ q_k^{(m)}=a\;\delta^k_0+m\;\delta^k_{m-1}
\end{equation}
so
\begin{equation}
\begin{aligned}
&p_0^{(m)}=b\ ;\ p_m^{(m)}=1\ ;\ q_0^{(m)}=a\ ;\ q_m^{(m)}=0\ ;\ \Delta_m=-a\\
&k\in[0,m-1]\ \begin{cases}
p_k^{(m-1)}=a^2\delta^k_0-mb\;\delta^k_{m-2}+a\;\delta^k_{m-1}\\
q_k^{(m-1)}=-a\delta^k_0-m\delta^k_{m-1}
\end{cases}
\end{aligned}
\end{equation}
then
\begin{equation}
\begin{aligned}
&p_0^{(m-1)}=a^2\ ;\ p_{m-1}^{(m-1)}=a\ ;\ q_0^{(m-1)}=-a\ ;\ q_{m-1}^{(m-1)}=-m\ ;\ \Delta_{m-1}=-(m-1)a^2\\
&k\in[0,m-2]\ \begin{cases}
p_k^{(m-2)}=m a b\;\delta^k_{m-3}+(m-1)a^2\;\delta^k_{m-2}\\
q_k^{(m-2)}=-(m-1)a^2\delta^k_0+m^2b\;\delta^k_{m-2}
\end{cases}
\end{aligned}
\end{equation}
this structure will repeat, indeed, if
\begin{equation}
k\in[0,m-j]\ \begin{cases}
p_k^{(m-j)}=A_j\;\delta^k_{m-j-1}+B_j\;\delta^k_{m-j}\\
q_k^{(m-j)}=-B_j\delta^k_0+C_j\;\delta^k_{m-j}
\end{cases}
\end{equation}
then $p_0^{(m-j)}=0$, $p_{m-j}^{(m-j)}=B_j$, $q_0^{(m-j)}=-B_j$, $q_{m-j}^{(m-j)}=C_j$, then $\Delta_{m-j}=B_j^2$ and the next coefficients are:
\begin{equation}
k\in[0,m-j-1]\ \begin{cases}
p_k^{(m-j-1)}=-B_jA_j\;\delta^k_{m-j-2}-B_j^2\;\delta^k_{m-j-1}\\
q_k^{(m-j-1)}=B_j^2\delta^k_0+C_j A_j\;\delta^k_{m-j-1}
\end{cases}
\end{equation}
so we have the recurrence $A_{j+1}=-B_j A_j$, $B_{j+1}=-B_j^2$ and $C_{j+1}=C_j A_j$ from $j=2$ (with $A_2=m a b$, $B_2=(m-1)a^2$, $C_2=m^2b$) up to $j=m-2$. At $j=m-1$ we arrive then to:
\begin{equation}
k\in[0,1]\ \begin{cases}
p_k^{(1)}=A_{m-1}\;\delta^k_{0}+B_{m-1}\;\delta^k_{1}\\
q_k^{(1)}=-B_{m-1}\delta^k_0+C_{m-1} \;\delta^k_{1}
\end{cases}
\end{equation}
with $p_0^{(1)}=A_{m-1}$, $p_1^{(1)}=B_{m-1}$, $q_1^{(1)}=C_{m-1}$ and $q_0^{(1)}=-B_{m-1}$ so $\Delta_1=C_{m-1}A_{m-1}+B_{m-1}^2$ and the last iteration gives the constant:
\begin{equation}
\begin{cases}
p_0^{(0)}=-B_{m-1}^2-A_{m-1}C_{m-1}\\
q_0^{(0)}=C_{m-1}A_{m-1}+B_{m-1}^2
\end{cases}
\end{equation}
the recurrence on $B_j$, $A_j$ and $C_j$ gives ($j\geq2$)
$$
B_j=-\bigl((m-1)a^2 \bigr)^{2^{j-2}}\ ;\ A_j=m\;a\;b\bigl((m-1)a^2 \bigr)^{-1+2^{j-2}}\ ;\ C_j=(m-1) m^j a^j b^{j-1} \left((m-1) a^2\right)^{2^{j-2}-j}
$$
so the final contant term is
$$
(m-1)^{-m+2^{m-2}+1} a^{2^{m-1}-m} \left(m^m b^{m-1}+(m-1)^{m-1} p^m\right)
$$
we can factorise the constant and the discriminant is then~\cite{BB}
\begin{equation}
m^m b^{m-1}+(m-1)^{m-1} a^m
\end{equation}

\section{Conclusions}
The algorithm developed here could be use for formal or numerical calculation of the {\tt gcd} of two polynomials, or the discriminant and the resultant. It doesn't use matrix manipulation nor determinant calculations and it takes $O(n)$ steps to achieve the goal. It provide also the two polynomials needed for  Bezout identity.

\appendix*
\section{1}
The Mathematica program for the algorithm is:

\begin{verbatim}
GCDList[{list1_, list2_, P_, Q_}] := {
Drop[First[list] list1 - First[list1] list2, 1], 
Drop[Last[list2] list1 - Last[list1] list2, -1], 
(First[list2] P - First[list1] Q)/x, 
 Last[list2] P - Last[list1] Q
 }
\end{verbatim}
this routine doesn't test the $\Delta_m$. The variable {\tt P} and {\tt Q} are there just for keeping track of the linear combination on {\tt P} and {\tt Q} which leads to the next step.

\begin{verbatim}
GCDListMax[{list1_, list2_, P_, Q_}] := 
 Module[{p1, q1}, 
  If[Last[list2] First[list1] - Last[list1] First[list2] == 0, 
   Return[{list1, RotateRight[First[list1] list2 - First[list2] list1], 
   P, x ( First[list1] Q - First[list2] P)}],
   p1 = Drop[First[list2] list1 - First[list1] list2, 1]; 
   q1 = Drop[Last[list2] list1 - Last[list1] list2, -1]; 
   Return[{p1/Max[p1], q1/Max[q1], 
     1/Max[p1]/x (First[list2] P - First[list1] Q), 
     1/Max[q1] (Last[list2] P - Last[list1] Q)}]]]
\end{verbatim}
this routine test the $\Delta_m$ and use the ``max" to normalise the coefficients at each step.

\bibliographystyle{plain}	% (uses file "plain.bst")
\bibliography{GCDPolynomial}		% expects file "myrefs.bib"

%\end{document}

\end{document}